\documentclass{llncs}

\usepackage{color}
\usepackage[cmex10]{amsmath}
\usepackage{amsfonts}
\usepackage{amssymb}
\usepackage[colorlinks=true,linkcolor=black,urlcolor={blue!50!black},citecolor=black]{hyperref}
\usepackage{tikz}
\usepackage[shortlabels]{enumitem}
\usepackage{multirow}
\usepackage{xcolor,colortbl}
\usepackage[title]{appendix}
\usepackage{capt-of}

\usepackage{soul}

\usepackage{csquotes}

\definecolor{darkgreen}{cmyk}{0.4,0.1,0.4,0.4}
\definecolor{gray}{cmyk}{0.3,0.1,0.1,0.2}

\newcommand{\blankline}{\vspace*\baselineskip}

\begin{document}

\title{Shepherd: Enabling Automatic and Large-Scale Login Security Studies}

\author{Hugo Jonker\inst{2,3} \and
	Jelmer Kalkman\inst{2} \and 
	Benjamin Krumnow\inst{1,2} \and 
	Marc Sleegers\inst{2} \and 
	Alan Verresen\inst{2}
	}
\authorrunning{Jonker et al.} 

\institute{Technische Hochschule K\"oln, Germany\\
	\email{benjamin.krumnow@th-koeln.de}, \url{http://www.th-koeln.de}
	\and
	Open Universiteit, Heerlen, Netherlands\\
	\email{hugo.jonker@ou.nl},
	\url{http://www.open.ou.nl/hjo/}
	\and
	iCIS Institute, Radboud University, Nijmegen, Netherlands
}

\pagestyle{plain}

\maketitle

\begin{abstract}
More and more parts of the internet are hidden behind a login
field. This poses a barrier to any study predicated on scanning the
internet. Moreover, the authentication process itself may be a weak
point. To study authentication weaknesses at scale, automated login
capabilities are needed.
In this work we introduce \emph{Shepherd}, a scanning framework to
automatically log in on websites. The Shepherd framework enables us to
perform large-scale scans of post-login aspects of websites. Shepherd
scans a website for login fields, attempts to submit credentials and
evaluates whether login was successful. We illustrate Shepherd's
capabilities by means of a scan for session hijacking susceptibility. 
In this study, we use a set of unverified website credentials, some of
which will be invalid. Using this set, Shepherd is able to fully
automatically log in and verify that it is indeed logged in on 6,273
unknown sites, or 12.4\% of the test set.
We found that from our (biased) test set, 2,579 sites, i.e., 41.4\%,
are vulnerable to  simple session hijacking attacks.
\end{abstract}

\section{Introduction}

As online services play an ever-increasing role in daily life, the
ability to ascertain privacy and security of these services grows
evermore important. Recent studies into privacy and security of websites
either scan up to the login point, or use manual logins. However, more
and more content is hidden behind logins. This causes two problems.
First and foremost, it hinders content shielded by a login from being
evaluated by such studies. Recent frameworks such as
FPDetective~\cite{AJNDGPP13} or OpenWPM~\cite{EN16} focus on studying
privacy aspects of websites. Such frameworks can be -- and have been --
used to study how websites handle various privacy aspects of their
visitors. However, to study the privacy of website \emph{users}, more is
needed: a scanner has to look beyond the login point. While this is
relatively easily done manually, or with user assistance, neither of
those approaches scale well. A second problem follows from the fact that
an authentication process is a security process -- and it may be
insecure. Indeed, the Open Web Application Security Project (OWASP) has
consistently ranked ``broken authentication and session management'' in
the top 3 security problems for websites since 2010~\cite{owasp17}. As
the number of sites with logins increase, the amount of sites with
problems in this category is likely to increase as well. Estimating how
many sites are insecure cannot be done manually -- there are simply too
many websites. Some form of automation is needed, yet there currently
are no tools to facilitate such studies. This is acknowledged as a
difficult problem. For example, Tang, Dautenhahn and King state that
``\emph{it is infeasible to evaluate cookie protection
automatically}''~\cite{TDK11}. Indeed, Qualys SSL Lab's SSL server test
may award top marks to a website that is susceptible to simple session
hijacking, which is clearly stipulated by Qualys.

There are two approaches to circumventing this problem: using manual
intervention or piggy-backing on single sign-on frameworks. However,
each of these has its limitations: manual intervention does not scale
well, and logging in with single sign-on frameworks often requires a
`registration' step on the site in question, and is therefore not akin
to an actual login.

We present \emph{Shepherd}, a scanning framework that is able to guide
(or ``shepherd'') scans beyond a login barrier. Firstly, Shepherd
enables a sorely needed automated, large-scale evaluation of
authentication process security. But secondly, and more importantly, it
also enables any other scan to be conducted on those parts of the
internet for which a login is required.

\subsubsection{Contributions.}
This work presents the following contributions:
\begin{itemize}
\item We design and develop \emph{Shepherd}, a tool that can perform
	large-scale scans of websites that require logins. This opens up
	the possibility to perform security scans on websites beyond
	the login barrier.
\item We illustrate Shepherd's capabilities by studying session
	hijacking susceptibility, with two sources for login
	credentials: BugMeNot, an open (but biased) database of usernames and
	passwords, and Facebook Single Sign On. 
	\begin{itemize}
	\item We show an approach to probe a user's cookie jar for
		session cookies of targeted sites.
	\item Using Shepherd, we can scan an order of magnitude more
		sites than achieved in previous (typically
		manually-assisted) studies.
	\item We find that out of 6,273 sites (from BugMeNot) scanned
		for session hijacking susceptibility, at least 2,579
		sites (i.e., 41.4\%) are vulnerable.
	\item BugMeNot is a biased source of credentials (certain
		websites are excluded, and any website is removed upon
		request). We compare the results of the biased BugMeNot
		experiment with the results of the (smaller) Facebook
		experiment.
	\end{itemize}
\end{itemize}

\subsubsection{Ethical considerations.}
Executing this study leads to several ethical concerns. First of all, we
submitted our study to our Institutional Review Board for Computer
Science and incorporated their advice in our studies. Secondly, we faced
the question of how to acquire a large set of login credentials from a
legitimate source. Fortunately, the BugMeNot database is exactly this: a
large set of login credentials with strict (and enforced) policies to
ban a site from inclusion upon request of the site owner. Thirdly, the
experiments must not exceed their mandate and break things. To this end,
the experiments were confined to only execute a login. The security
evaluation then occurred on the client side. Thus, the connection was
broken following the response to a login attempt (irrespective of the
success of the attempt). We worked on this by testing Shepherd on a
small number of domains and resolving any issues. The results are not
100\% perfect, but the fraction of mistakingly pressed buttons we
detected is very small. Fourthly, the results could be used to create
attacks against specific websites. Therefore, we only discuss results in
aggregate, not for particular sites. Fourthly, the tools created can
easily be misused. Therefore, we will not publicly release them, but we
will make them available to other bona fide researchers and reviewers
upon request. Finally, we used responsible disclosure to notify sites we
manually found to be vulnerable.

\section{Related work}

Several studies have been performed that needed to authenticate with
websites. Typically, these studies leveraged manual intervention to
achieve logins. While this approach successfully accounts for the
post-login stage, the studies using this approach are limited in scale
to the participants and difficult to repeat.

Mundada et al.~\cite{MFK16} let participants log into a websites to
analyse security of the login process of 149 sites. They found several
security risks in well-known sites such as Yahoo. These findings
underscore the importance of extending such studies to more sites.
However, Mundada et al.~asked volunteers to log in, which is
labor-intensive.

Similarly, Wang et al.~\cite{WCW12} study the security of various
single sign-on implementations by manually logging in. They uncovered
various security flaws in widely-used single sign-on implementations.
This is especially troubling given that since Wang et al.~have performed
their study, single sign-on services have been upgraded and updated, and
typically are somewhat backwards compatible. Moreover, their study
focuses on single sign-on service providers popular in the US, but does
not account for single sign-on service providers from
elsewhere\footnote{E.g., the service by
Vkontakte, a Facebook alternative popular in Russia.}.

A different approach to logging in was taken by Kranch and
Bonneau~\cite{KB15}. They studied the use of
HSTS\footnote{\emph{Hypertext Strict Transport Security}, telling the
browser to only connect on future visits using HTTPS.} and
HKPK\footnote{\emph{HTTP Public Key Pinning}, telling browsers to keep
the server's public key for future visits.}, scanning sites while being
logged in with multiple single sign-on providers. They achieved this by
keeping session cookies in the cookie jar. While this approach aims to
study impacts of single sign-on providers on websites, it does not enable
access to post-login areas.

To our best knowledge Robinson and Bonneau~\cite{RB14} provide the only
work which attempts to automatically login into websites via single sign-on
providers. However, the authors manually selected websites with Facebook
Connect. Moreover, their study focused on what access to the user's
Facebook profile these sites obtain after logging in. Conversely, our
work focuses not on the single sign-on provider, but on the logged-in
site.

\section{High-level overview of Shepherd}
\label{sec:high-level_overview}

\begin{figure}[t]
\centering
\includegraphics[width=.9\textwidth]{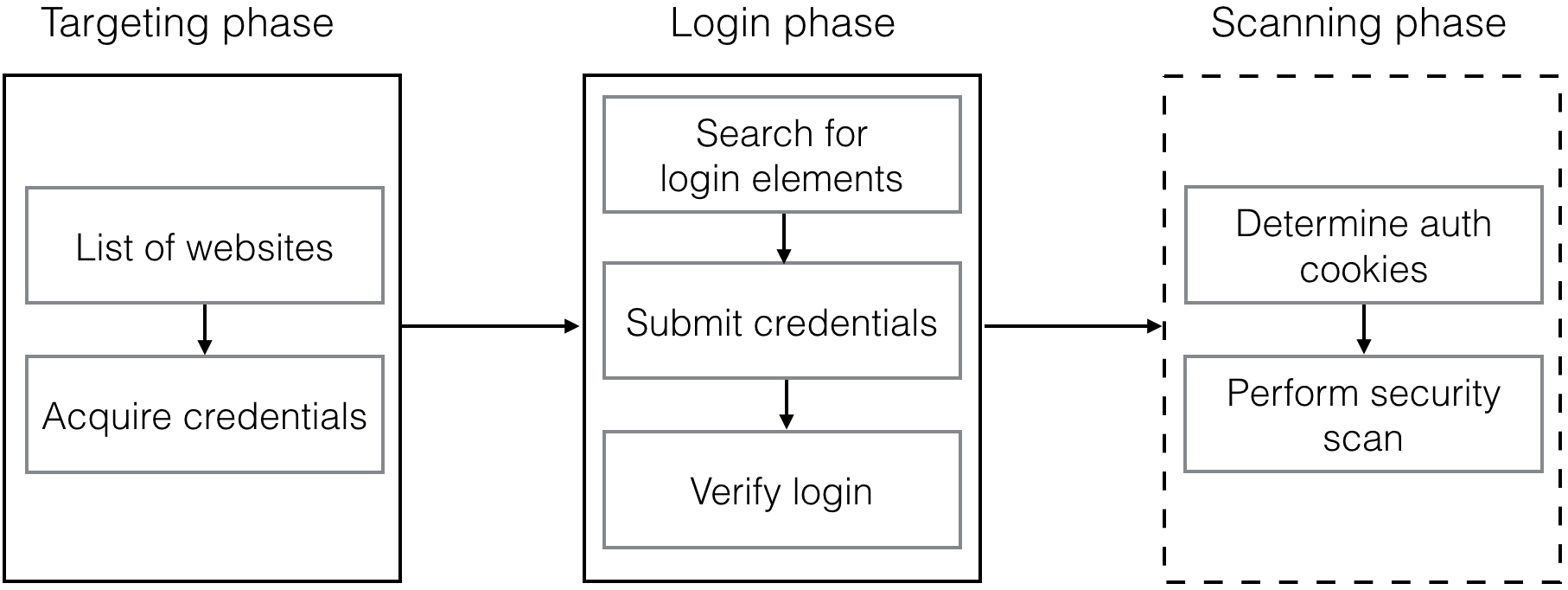}
\caption{Overview of Shepherd's design (scanning phase shows the design of
the experiment from Section~\ref{sec:case_study}.}
\label{fig:design}
\end{figure}

In this section, we present a high-level overview of the design of
Shepherd, and discuss various challenges and design decisions.
The high-level structure of Shepherd is depicted in
Fig.~\ref{fig:design}. Shepherd consists of three phases:
\begin{enumerate}[I.]
\item targeting phase, selecting targets of the scan and obtaining
	credentials;
\item login phase, which performs the logins and evaluates their
	success;
\item scanning phase, which executes the desired scan.
\end{enumerate}

The targeting phase may be executed in a stand-alone fashion, or on the
fly. In early testing, we found that on-the-fly targeting introduces
bottlenecks, which are irrelevant to the actual scan. As such, we
chose to redesign this phase to operate stand-alone, independent of the
rest of the framework. As such, the results of the targeting phase are
stored in a database, which can be read at any time by the subsequent
phases.

Conversely, the actual evaluation of the scan may be executed on the
fly, or \emph{a posteriori}, by examining the data collected on the fly.
In either case, some data needs to be processed, so at least part of the
scanning phase (data collection) must operate on the fly.

The login phase is by far the most complex part of Shepherd, and merits
separate discussion. This phase is discussed in detail in
Sec.~\ref{sec:login_phase}. Noteworthy details pertaining to the other
phases are discussed below.

\subsection{Preparation phase}
\label{sec:preparation_phases}
Shepherd uses a list of target websites and corresponding credentials as
input in order to login. The purpose of the preparation phase is to
automatically obtain the credentials given the list of target websites.
There are two options to obtain credentials automatically: registering
new users (and thereby creating new credentials), or relying on existing
databases of credentials.

Each of these approaches has drawbacks. A major drawback to
automatically creating accounts is that typical account creation
processes guard against precisely such actions (e.g.~by means of
captcha's). This could be circumvented by manual creation, which is
laborious and does not scale well. Conversely, while obtaining credentials
from an existing database of credentials is more straightforward, this
too has its downsides. A major methodological issue is that the
selection of targeted websites is limited by the pre-existing database.
A more practical consideration concerns the legitimacy of the database.
We found various online services containing less-than-legally obtained
credentials, e.g.~passwords obtained from data breaches. 

Shepherd's targeting phase outputs credentials in a database. This
allows alternative targeting approaches (e.g.~manually-supplied
credentials) in addition to automatically sourced credentials. For
Shepherd's current targeting phase, we source credentials from the
BugMeNot website. BugMeNot provides logins to websites for users who do
not wish ``to be bugged''. To the best of our understanding, they
provide a legitimate service, with restrictions on sites added: no
pay-per-view sites, no banks, no age-restricted sites, and no sites
whose content is fully accessible without login. While this allows us a
rich source of credentials, it introduces also two issues. Firstly, the
validity of the credentials is uncertain. Not only can credentials be
invalid, BugMeNot can even have credentials for sites that do not have
logins. Secondly, it introduces a bias in the sites studied. As
mentioned, BugMeNot excludes several categories of sites, such as banks.
This has to be taken into account when examining the results of any
study.

\subsection{Scanning phase}
After logging in, the scan of the target website is executed. Shepherd
is written in Python, therefore, scans may be expressed as modules in
Python. Shepherd provides an interface to interact with the browser and
detect effects of interactions. This interface is a wrapper of Selenium
commands, but streamlines error handling and ensures
performance-optimised commands are used by the scanning module.
Optionally, multiple modules can be hooked into Shepherd. This allows
for sequential execution of several scanning modules. Scan results are
determined on the fly and stored in CSV files for a posteriori
evaluation.

By default, Shepherd determines which cookies are authentication cookies
and performs a security scan. However, it is possible to add custom
interactions with the browser. Shepherd provides an interface which
functions as a wrapper for Selenium commands. The results of the scan
are stored in CSV files, which can be externally processed for a
posteriori evaluation. 

\subsection{Implementation}
Shepherd uses Selenium, a testing tool for automating interaction with
browsers. According to Englehardt and Narayanan~\cite{EN16}, headless
browsers are sometimes served different content than full browsers.
Moreover, they also found that in headless browsers, a substantial
number of sites render incorrectly or crash due to missing plugins. To
avoid these drawbacks, Shepherd uses a full browser for scanning.
Shepherd can use either Chrome or Firefox as a designated browser, and
runs on Linux and on MacOS systems.

\subsubsection{Performance.}
\label{performance}
Logins are typically slow and can easily take several seconds.
When attempting to login on unknown sites, using a form which may or may
not be the login form, with credentials that may or may not be valid,
several passes have to be taken. To study sites at scale with all these
factors in mind thus necessitates high performance. 

With respect to optimisation, we found that sometimes it pays to replace
Selenium built-in functions with our own JavaScript code. Accessing the
plain HTML content of elements takes 14 ms with Selenium. This results
in high runtime, if a site uses a large number of elements. In one
example, accessing and filtering elements took 16.8~sec compared to
50~msec with Javascript. Another example is Selenium's function to query
for multiple elements, which takes 1~sec per query. This sums quickly
up, when executed multiple times, due to multiple windows and iframes.
With Javascript the same operation plus applying filters to all elements
can be completed in 200~msec.

\subsubsection{Stability.}
Shepherd's stability is of course important when performing automated
scans. There are several caveats that we encountered. First,
some websites overwrite basic Javascript functions. This may cause
problems, which is why Shepherd executes its Javascript in a separate
window. 

On the level of running processes, we found that sometimes, a browser
instance crashes or runs out of memory. To mitigate out of memory
errors, Shepherd executes browser instances as threads. If the instance
exceeds memory limits, Shepherd kills the thread and starts a new
instance. To mitigate crashes, browser instances are restarted after a
number of runs. With these measures in place, we find Shepherd can scan
and login to about 1500 sites per browser instance per day.

\subsubsection{Single sign-on logins.}
\label{sec:facebook_implementation}
We experienced single sign-on logins as less effective than hoped. Typically,
following authentication via single sign-on, the website requests that the user
fills in a registration form of some kind. Access to other parts of the
site is typically blocked until this form is filled in. This behaviour
does not always occur, but frequently enough that it affects the study.
This implies that single sign-on logins by themselves are insufficient
to study the post-login aspects of websites. 
Nevertheless, single sign-on logins are effective in studying security
of the login process itself.

To that end, we extended Shepherd with limited capabilities for using
Facebook's single sign-on service. This capability begins with searching
for a Facebook login entry point on the website, which need not be
present (in contrast to~\cite{RB14}). Once Shepherd believes it has
found such an entry point, it processes the login as a regular login,
but with the loaded Facebook credentials.

\section{Login Phase Design} \label{sec:login_phase}
The design of the login phase is to consider the target website as a
black box. That is: it cannot incorporate any knowledge about the site
design or layout except that which it learns by scanning the site.
Therefore, we device a generic model of the website login process.
This model provides a three-step approach to logging in:
\begin{enumerate}
\item Login page detection: in the first step, the login page of the site is
	identified. This step is concluded successfully once the login
	area is identified.
\item Authentication: in the second step, credentials are submitted. This step
	is concluded successfully if a change in the website is detected
	following submission of credentials.
\item Verification: the last step verifies whether the website is indeed
	logged in. Whereas the authentication step uses a crude measure
	(change in webpage), this step uses an array of more
	sophisticated techniques.
\end{enumerate}
Note that for a fully automated approach, false positives and false
negatives may crop up in each of these steps. We evaluate the
performance of Shepherd in this respect in
Section~\ref{sec:scanner_performance}.

\begin{figure}[th]
	\hspace{-1.2cm}
	\includegraphics[width=1.2\textwidth]{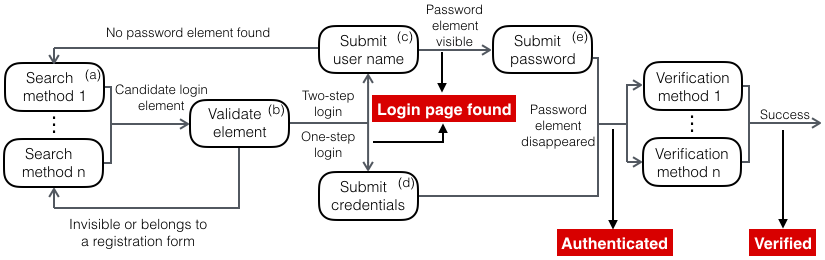}
	\caption{Overview of the login process.}
	\label{fig:login_scheme}
\end{figure}

Each of these steps consists of numerous actions, leading to the model
depicted in Figure~\ref{fig:login_scheme}. In this figure, the red boxes
indicate successful conclusion of the corresponding step.

\subsection{Searching for login elements}
As shown in Figure~\ref{fig:login_scheme}, different methods for finding
login elements may be used (a). Shepherd uses
the following five methods to find a login page. Starting from the
landing page, Shepherd first looks for
suitable login elements (method~1). Failing that, Shepherd selects
elements with relations to login-based terms from the landing page.
Selected urls are visited first (method~2). All links on the visited
pages are cached for method~5. Should this step also fail, clickable
elements are clicked (method~3). Failing this, Shepherd attempts a brute
force search of the login page by appending login related
keywords\footnote{Shepherd contains a dictionary with multiple
translations for keywords from native speakers and Google translations.}
to the base
url\footnote{Specifically: http://target/login, http://target/account
and http://target/signin.} (method~4). In case this also fails, the last
approach (method~5) is to follow all
links collected during method~2. This approach accounts for sites that
require a user to follow more than one link before being able to enter a
username, such as imdb.com.

After each action in finding login elements, Shepherd scans all open
windows and iframes looking for a password field (cf.~`b' in
Figure~\ref{fig:login_scheme}). When Shepherd encounters a visible input
element of type password which is not part of a registration form, it
assigns the status \textit{login page found}. A form is considered a
registration form if it contains more than 3 visible input elements
(including the found password field).

If a candidate login page is found, Shepherd moves on to submitting
credentials. Two types of credential submission processes are
considered: one-step logins, in which a username and a password are
submitted in one go, and two-step logins, in which the user needs to
press a ``submit'' button after entering the username. One-step login
candidates are easily detected by password elements. Conversely,
two-step login elements must be distinguished from other input elements
(e.g.~search fields or newsletter sign-up forms), and can only be
detected after a valid username is submitted and the password field is
shown. Therefore, two-step logins can only be successfully
black-box detected using a valid username.

\subsection{Submitting credentials to login}
After identifying a login area, Shepherd proceeds to submit credentials
according to the type of login area (one-step or two-step). If this
causes the password input element to disappear, the website status is
set to \emph{authenticated}. This heuristic is imperfect, as websites
may also stop showing password fields in other cases. For example, when
the user is blocked or a server error results in a 404 page. To separate
such cases from actual logins, the authentication process is followed by
a verification process.

\subsection{Verifying login status}\label{sec:verifier}
After submitting credentials, Shepherd evaluates whether the login was
successful. Shepherd's approach to this is similar to previous
approaches~\cite{MFK16,CTBO14}. The main difference is that previous
works relied on manual logins, and were therefore certain of the
validity of the credentials. Conversely, Shepherd relies on a source of
unverified credentials, and thus cannot assume that the credentials are
correct. To improve verification, each verification method is run twice:
once with all cookies, and once without any cookies. The test is passed
when the first run succeeds \emph{and} the second run fails. This
ensures that the verification method only depends on the site's cookies.
Currently, Shepherd uses two verification methods. The first is to
detect a logout button or user identifier on the landing page. The
second is to attempt to re-open the login area. This is typically not
possible in well-designed sites after the user has logged in.

\section{Scanning for session hijacking susceptibility}

To validate the effectiveness of Shepherd, we set out to conduct a
large-scale experiment to ascertain how many sites are vulnerable to
session hijacking. Various previous works have investigated session
security, e.g.~\cite{MFK16,AHS17}. However, no study attempted
large-scale, fully automatic login. To automate the scan, we simplified
the goal and only look for authentication cookies for which the
\texttt{Secure} flag is not set. This is sufficient, in case these
cookies can indeed be stolen and if these cookies are sufficient for
session hijacking. We performed two small experiments to test if these
assumptions hold in practice.

An easy way to determine session hijacking susceptibility is to actually
hijack the session. In controlled conditions (at the very least: the
researcher's own session is hijacked on another machine, and nothing is
done with the session except confirming it is hijacked), a manual attack
could be performed ethically. However, it is not clear whether it is
possible to automate this approach yet still maintain ethical
safeguards. For example, how to ensure that nothing is clicked after
logging in, if we cannot be 100\% sure that the browser is logged in?

To sidestep this issue, we first investigate whether the question can be
simplified, based on the following assumptions:
\begin{enumerate}[I.]
\item Possession of a victim's authentication cookie is sufficient to
	execute a session hijacking attack;
\item Lack of the cookie flag \texttt{Secure} is sufficient to steal a
	victim's cookie.
\end{enumerate}
These assumptions are well-known\footnote{Assumption I was even used to
automate session hijacking, e.g. by the Firesheep plugin for Firefox.}
and believed to hold. However, in practice, other security practices may
interfere with these. For example, the \emph{Hypertext
Strict Transport Security} HTTP header instructs browsers to only
contact a site over an HTTPS channel, and never over an HTTP channel.
Thus, lack of a Secure cookie-flag may still be mitigated elsewhere.
Similarly, an authentication cookie may be bound to e.g.~the IP address of a
computer. While this countermeasure is far from perfect, it would defeat
assumption~I.

\subsubsection{Cookie stealing suffices for session hijacking.}
We manually tested a handful of sites found in an early trial run of
Shepherd. For each tested site, we found that its authentication cookie 
sufficed to perform a session hijacking attack. Though this was a
limited test, it always succeeded. This gives a reasonable degree of
assurance that in practice, cookie stealing suffices for session
hijacking.
All sites investigated in this part of the experiment were sent notice
of this, though most did not react or acknowledge reception.

\subsubsection{Automatically probing a cookie jar.}

We devised a means to automatically probe a victim's cookie jar for
authentication cookies. Similar to Van Acker et al.~\cite{AHS17}, we use
ARP poisoning and the MitMproxy (\url{http://mitmproxy.org/}) to set up
a man-in-the-middle. Our man-in-the-middle injects HTML into any
unencrypted HTTP response (i.e., not transmitted over HTTPS).
Specifically, for each target site,
it injects a line into the HTML header as follows:

\verb+<link src="http://TARGETSITE/style.css" type="text/css">+

\noindent
By using links to style sheets, we avoid common measures that would
interfere, such as blocking third party scripts.

This line triggers the browser to send out an HTTP request (i.e.,
unencrypted) to the target site. Due to the design of HTTP, cookies are
automatically included unless settings (such as the \texttt{Secure}
cookie flag) prevent this. Thus, the injection will trigger the victim's
browser to send out an HTTP request to each targeted site, accompanied
by session cookies if these are present. Testing this setup with a
handful of sites showed that indeed the cookie jar can be easily probed
and authentication cookies can be stolen.

\blankline \noindent
From these two experiments we conclude that examining the \texttt{Secure}
flag on the authentication cookies is indeed sufficient in practice to
determine whether a site is susceptible to session hijacking.
In order to use Shepherd to scan for authentication cookies lacking the
\texttt{Secure} flag, we must be able to automatically
determine which of the cookies are authentication cookies.

\subsection{Determining authentication cookies}
Previous work attempted to identify authentication cookies by applying
heuristics on cookie properties \cite{RNDPJ12,TDK11,BCFK14,NMYJJ11}.
Calzavara et al.~\cite{CTBO14} found that these approaches either
resulted in an over- or under-approximation of which cookies are
authentication cookies. They provide an improved algorithm, which is
used by Mundada et al.~\cite{MFK16} and which forms the basis of our
approach. This approach eliminates cookies by assessing the user's login
status. As such, a reliable verification of this status is a necessary
pre-condition. The approach is to remove a cookie from the set of
cookies and access the site afresh. If the user is then still logged in,
the removed cookie was not an authentication cookie and is removed from
consideration. As websites may use multiple cookies for authentication,
applying this test means testing combinations of cookies. The number of
tests to check all combinations quickly becomes infeasible in practice,
especially for large-scale measurements. Our current implementation
adapts two out of three measures from~\cite{CTBO14} to reduce the total
number of needed site revisits. Once a working set is found, all
supersets can be excluded from further tests, since these sets will work
as well. Additionally, if a not working set is found, all subsets can
also be eliminated, as these lack at least one cookie needed to
authenticate. We measured the number of executions and runtime of this
algorithm in our study. The results for performing this algorithm can be
found in Section~\ref{sec:case_study}.  

\section{Case studies: session hijacking susceptibility}
\label{sec:case_study}

We tested Shepherd's ability to perform large-scale, post-login scanning
of websites by evaluating session hijacking susceptibility. We scanned
over 50,000 websites using credentials from BugMeNot. In addition, we
scanned the Alexa Top 10,000 with an extension to login using Facebook's
single sign-on solution.

We performed the BugMeNot study in February and the Facebook study in
April 2018. We used two machines hosted in Germany, each running five
browser instances in parallel. Shepherd needs an average of 60.1
seconds to process a single site based on our scan with credentials from
BugMeNot. This allowed us to process 15,000 sites a day. The Facebook
extension can scan 3,000 sites per day. This lower limit is due to
websites that heavily use Facebook-related elements. Both scans were
completed within one week.

We also measured Shepherd's performance in determining authentication
cookies. In order to keep the runtime acceptable, we did not execute
this process for websites with more than 23 cookies. In our study,
websites contained an average of 8.2 cookies. Shepherd needed (on
average) 11.9 revisits to determine which were authentication cookies.
On average, websites in our experiment use 1.51 cookies for
authentication, which is close to the observation in~\cite{CTBO14}.

\subsection{Extraction of  BugMeNot credentials}

\begin{figure}[th]
\centering
\includegraphics[width=.6\textwidth]{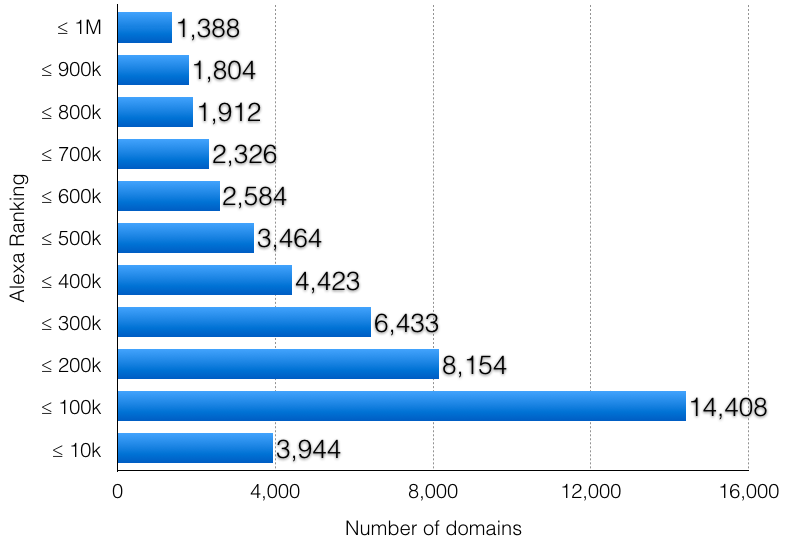}
\caption{Distribution of the BugMeNot dataset within the Alexa Top
	1 Million.}
\label{fig:bmn_distribution}
\end{figure}

For our case study, we sourced credentials from BugMeNot by applying
domains from the Alexa Top 1 Million dataset from February 2018. With
this approach we extracted 131,345 accounts for 50,439 unique domains.
As shown in Figure~\ref{fig:bmn_distribution}, this dataset covers over
39\% of the Alexa Top 10K domains and 18\% of the Top 100K respectively.
After consulting the IRB, we discarded our attempt to extended this set
of credentials with credentials from a prior  run in October 2017. The
major reason for that is, that this set might contain websites, which
have decided to opt-out of BugMeNot.  

\subsection{Scanner Success Rate}
\label{sec:scanner_performance}
Table~\ref{tbl:scanner_performance:all} shows success rates of the
various processes (described in Section~\ref{sec:login_phase}) Shepherd
executes to perform logins. As previously mentioned, the underlying
heuristics are not 100\% perfect and may occasionally fail to determine
the status correctly. To determine bounds on the error rates, we
manually evaluated websites previously scanned by Shepherd. This
revealed shortcomings and areas for improvement of Shepherd's approach
to logging in. To evaluate Shepherd, we created four sets of 100
websites each from the BugMeNot case study. In the first case, we
manually accessed sites, while for cases~2--4 we reviewed automatically
created screenshots.

\begin{table}[t]
	\begin{tabular}{p{3.5cm}|rp{0.2cm}rr|p{0.5cm}|rp{0.2cm}rr|}
		 \multicolumn{1}{c}{}& \multicolumn{4}{c}{BugMeNot scan}
		 & \multicolumn{1}{c}{} & \multicolumn{4}{c}{Facebook scan} \\
		 \cline{2-5}\cline{7-10}
		 & \# sites && out of &&&                    \# sites && out of &\\
		\cline{2-5}\cline{7-10}
		Total			& 50,439 && --      &          && 10,000 && -- & \\
		Sites reached 		& 46,942 &&  50,439 &  (93.1\%) &&  8,829 && 10,000 & (88.3\%) \\
		Login page found$^1$& 31,543 &&  46,942 & (67.1\%) &&  2,057 &&  8,829 & (23.3\%) \\
		Authenticated$^2$	& 9,903 &&  31,543 & (31.4\%) &&  1,915 &&  2,057 & (93.1\%) \\
		Verified$^{3,4}$	&  6,273 &&  9,903 & (63.3\%) &&    383 &&  1,915 & (20.0\%) \\
		Auth cookies found	&  5,539 &&   6,273 & (88.3\%) &&    330 &&    383 & (86.2\%) \\
		\cline{2-5}\cline{7-10}
	\end{tabular} \\[1.5ex]
	$^{1,2,3,4}$ Success rate analysed in corresponding case number.
	\vspace{1.5ex}
	\caption{Performance of the BugMeNot and Facebook scans.}
	\label{tbl:scanner_performance:all}
\end{table}

\begin{description}
\item[Case 1: Finding login pages.]
	This case concerns websites where Shepherd was unable to
	find login areas. We evaluated 100 such sites found login areas on
	35 sites. 15 sites could not be evaluated, because they were
	blocked for foreign countries or did not respond. We did not
	discover login elements on the other 50 sites. Therefore,
	Shepherd missed out on at most 35 out of 85 sites. While this
	seems reasonable, this concerns the first stage. Therefore, any
	improvements in the success rate here will translate into
	significantly more sites to process further.

\item[Case 2: Authentication failures.]
	This case concerns sites where Shepherd found a login area and
	submitted credentials, but could not detect success. Shepherd
	automatically took screenshots of 100 such sites.
	We found 55 screenshots indicating incorrect credentials. 30
	pages could not be evaluated due to unclear
	content (23), missing screenshots (6) or inappropriate content.
	6 screenshots showed dialogues for captchas and 3 showed 404 error
	messages. Only on 6 screenshots we saw that Shepherd ended up in
	a wrong login area (SSO logins: 2, registration pages: 2, wrong
	forms: 2). Therefore, we conclude that given a login area, the
	process of credential submission and evaluation of the success
	of that step performs reasonably well. However, 55 out of 70
	sites showed errors due to invalid credentials. Finding a better
	service with a massive amount of credentials is in our view
	unlikely\footnote{In our search, we found a handful of 
	databases of dubious legality, and others that only included a
	subset of the credentials of BugMeNot}.
	The most obvious improvement approach is then to use single
	sign-on credentials.

\item[Case 3: Authenticated, but not verified.]
	This case concerns sites that successfully pass authentication,
	but verification of that failed. Verification is supposed to
	fail when login was not successful. In 60 out of 100 cases,
	this was the case, underscoring the need to verify whether login
	is indeed successful. 24~cases showed clear signs that Shepherd
	entered the post-login stage, while 16 cases could not be
	verified. In other words: in at least 24\% of the examined
	cases, this process resulted in a false negative. In the
	BugMeNot scan, 4,488 sites fell in this category. Additional or
	improved verification methods thus may lead to hundreds of
	sites more evaluated.

\item[Case 4: Verified.]
	This case concerns 100 sites which passed verification.
	Of these, only two sites could not be checked. Two other
	screenshots showed that the user account was banned. Nevertheless,
	Shepherd was clearly logged into these sites. For that, we
	find the verification process to have high accuracy ($\geq$96\%)
	and therefore have high confidence in all findings on
	sites marked \textit{verified}.

\end{description}

In addition, we compare votes from BugMeNot to our success rates in
Appendix~\ref{sec:correlation_votes}. This provides another approach to
assess Shepherd's effectiveness.

We zoomed in on the process to determine the login area. We tracked how
successful each of the methods used in this process were. Here, one
should keep in mind that these methods are executed sequentially: only
if methods 1--4 fail, method 5 is executed. Of the 31,543 domains where
a login area was found, method 2 was most successful, finding over half
the login areas.
\begin{itemize}
	\item Method 1 - Landing page: 6,865
	\item Method 2 - URLs: 17,381
	\item Method 3 - Clicking elements: 3,495
	\item Method 4 - Brute force: 3,126
	\item Method 5 - Second level URLs: 676
\end{itemize}

Overall, we recorded 6,273 successful and verified logins. For 3,630
sites, Shepherd believed it was logged in but failed to verify this.
This set may also contain sites where login was indeed successful. In
our manual evaluation, the rate where logins were successful for
authenticated-but-not-verified sites was 24\%. Extrapolating from that
experiment to this set gives a rough estimate of about 870 sites where
login was indeed successful, despite lack of automated verification.

\subsubsection{Performance of Facebook login extension.}
For this case study, we divided the Alexa Top 10K into 4 equal sets. In
addition, we created two Facebook accounts, one for each machine. The
first machine was used to scan the first 3 sets (see top part in
table~\ref{tbl:scanner_performance:all}), while the second machine
scanned only the last set. We observed that the Facebook account for the
first machine was blocked at a certain point while scanning the second
set. This was due to posting inappropriate content, caused by Shepherd
clicking share buttons on visited sites. We adjusted Shepherd to avoid
posting content by blacklisting certain types of Facebook URLs. After
recovering the blocked account and scanning the third set, we found 55
shared posts on the account (each of which must be due to a successful
login). Shepherd misclassified these as logins. Furthermore, we found
that 664 third parties were granted access to the used Facebook account.
These must originate from one of the 1,915 successful logins. The second
Facebook account was also blocked, this time due to suspicious
behaviour. Recovery of this account was more involved and therefore
omitted. Unlike the other 3 sets, we thus could not verify Shepherd's
results for this set in the Facebook account.

\begin{table}
	\centering
	\begin{tabular}{p{2cm}|cc|cc}
		& Auth.~but not verif. &\% & Verified & \% \\
		\hline
		Total	& 3,630  & --~& 6,273 & --~\\
		\hline
		HSTS &412 & 11.4\% & 905 & 14.4\% \\
		HKPK & 37 & 1.0\% & 33 &0.5\%\\ 
		\hline
		HttpOnly & 2,115 & 58.3\%	& 3,605 & 57.5\%\\
		Secure &  1,221	& 33.6\%	& 3,613 & 57.6\%\\
		HSTS/secure & 125 & 3.4\%  & 81 & 1.3\%\\
		SameSite & 0 & 0.0\%	& 0	& 0.0\%\\
		\textbf{Susceptible}  & \textbf{2,284}  & \textbf{62.9\%}
			& \textbf{2,579} & \textbf{41.4\%}\\
	\end{tabular}
	\vspace{1.5ex}
	\caption{Security evaluation of the entire BugMeNot data set.}
	\label{tbl:security_scan_bmn}
\end{table}

\begin{table}
	\centering
	\begin{tabular}{p{2cm}|cc|cc}
		& Auth. but not verif.  &\% & Verified & \% \\
		\hline
		Total	& 1,532  & --~& 383 & --~\\
		\hline
		HSTS & 391 & 25.5\% & 90 & 23.5\% \\
		HKPK & 7 & 0.5\% & 3 & 0.8\%\\ 
		\hline
		HTTPonly & 995 & 64.9\%	& 333 & 86.9\%\\
		Secure &  1,015	& 66.3\%	& 363	& 94,8\%\\
		HSTS/secure & 105 & 6.9\%  & 2 & 0.5\%\\
		SameSite & 0 & 0.0\%	& 0	& 0.0\%\\
		\textbf{Susceptible}  & \textbf{412}  & \textbf{26.9\%}	& \textbf{18} & \textbf{4.7\%}\\
	\end{tabular}
	\vspace{1.5ex}
	\caption{Security evaluation of the Facebook case study.}
	\label{tbl:security_scan_fb}
\end{table}

\section{Analysis of results}
\label{sec:security_evaluation}

Shepherd automatically assesses the security of websites by scanning
HTTP headers of a website's response and cookie flags. Assessing sites
without logging in may lead to incorrect conclusions about a website's
security. Therefore, only authenticated and verified sites are
considered for assessments. Note that based on our manual validation from 
Section~\ref{sec:scanner_performance}, only verified sites were
evaluated in a post-login stage with a high confidentiality. In case the
authentication cookies could not be determined, all cookies were treated
as authentication cookies. 

A website is marked as susceptible to session hijacking, if at least one
of the following three conditions is true:
\begin{enumerate}[I.]
\item the website does not set the \texttt{Secure} cookie flag, 
\item the website does not set the \texttt{Samesite} cookie flag, or
\item the website does not use HSTS.
\end{enumerate}

For verified sites, we find that 2,579 of 6,273 (41.4\%) sites are
susceptible to session hijacking. A smaller fraction of websites uses
HSTS. Of these, 81 sites use HSTS but do not set the secure flag. We
also found that the \texttt{HttpOnly} cookie flag was missing on 2,668
(42.5\%) websites. For sites scanned with Facebook credentials, we
discovered that 18 out of 383 sites (4.7\%) are susceptible to session
hijacking. Moreover, in the Facebook scan, 50 sites lacked the
\texttt{HttpOnly} flag (13.1\%). 

Tables~\ref{tbl:security_scan_bmn} and~\ref{tbl:security_scan_fb} show
the results of the security evaluation. All sites which were
authenticated, but not verified, are summarised in the category
authenticated. Sites are only denoted as susceptible when they use none
of HSTS, \texttt{Samesite} flag or \texttt{Secure} flag. Sites that do
not implement the \texttt{Secure} flag but use HSTS are counted under
HSTS/secure.

\section{Investigating the BugMeNot Bias}
\label{sec:comparison}

A key feature of the main experiment is that it uses a source of
usernames and passwords. However, that source, BugMeNot, has strict
rules on what is allowed -- for example, credentials from hacking
attempts are disallowed. Moreover, site owners can request BugMeNot to never
accept credentials for their sites. Thus, sites in BugMeNot are
not representative of the whole internet.

To get a better grasp on how biased the BugMeNot dataset is, we
compared the results of that experiment with the Facebook experiment. Of
course, the latter is also biased (not every site with a login will have
a Facebook login), but we assume that these biases are independent of
each other. In the comparison, we only use domains out of the Alexa Top
10K, as the Facebook experiment was confined to these. The results are
shown in Table~\ref{tbl:real-security_scan_fb_and_bmn}. In this table,
the percentages columns compares entries on that row to the row labeled
``total''.

\begin{table}[t]
\centering
\begin{tabular}{p{2cm}|c|cccc|c|c|cccc|}
\multicolumn{1}{c}{}	& \multicolumn{5}{c}{Facebook experiment} &
	\multicolumn{1}{c}{} & \multicolumn{5}{c}{BugMeNot experiment}\\
	\cline{2-6}\cline{8-12}
\multicolumn{1}{c|}{}	& Auth. & \multicolumn{4}{c|}{Verification\ldots} &
	& Auth. & \multicolumn{4}{c|}{Verification\ldots}\\
 & OK & ~failed & \%  & success &\% & & OK & ~failed & \%  & success &\% \\
\cline{2-6}\cline{8-12}
Total	& 1,915 & 1,532  & --~& 383 & --~ && 778 & 289  & --~& 489 & --~\\
\cline{2-6}\cline{8-12}
Secure & 1,378 & 1,015 & 66.3\% & 363 & 94.8\% && 512 & 147 & 50.9\% & 365 & 74.6\%\\
HSTS/secure & 107 & 105 & 6.9\%  & 2 & 0.5\% && 17 & 13 & 4.5\%  & 4 & 0.8\%\\
HTTPonly & 1,328 & 995 & 64.9\%	& 333 & 86.9\% && 459 & 189 & 65.4\% & 270 & 55.2\%\\
\textbf{Susceptible}  & \textbf{430} & \textbf{412}  & \textbf{26.9} \%	& \textbf{18} & 
\textbf{4.7}\% && \textbf{251} & \textbf{130}  & \textbf{45.0} \%	& \textbf{121} & 
\textbf{24.7}\%\\
\cline{2-6}\cline{8-12}
\end{tabular}
\vspace{1.5ex}
\caption{Comparison of successfully authenticated sites of two experiments}
\label{tbl:real-security_scan_fb_and_bmn}
\end{table}

From this comparison, it is clear that the BugMeNot approach is better
engineered: when it authenticates, verification is far more often
successful. Furthermore, differences between cookie flags are profound:
\texttt{Secure} flag differs by 20\%, the \texttt{HttpOnly} flag by
30\%. In the BugMeNot experiment, 24.7\% of websites where the login was
successfully verified, was susceptible to simple session hijacking.
However, in the Facebook experiment, this was only 4.7\%. Based on this
comparison, it is clear that the results of the BugMeNot experiment
should not be extrapolated as-is. At the same time, in both experiments,
insecure cookies are a common occurrence (at least 24.7\% and 4.7\%,
respectively). We therefore conclude that more effort is needed to
secure online authentication processes.

\section{Conclusions and Future work}
Many previous works have studied the web. Most of these could not
look beyond the login barrier on websites. However, logins have become
more common. This implies that more and more content that is actually
accessed by users is omitted from such studies. Research that attempted
to address the post-login world, mostly fell back on manual
intervention, to avoid the many challenges with an automatic
approach~\cite{TDK11}. 

In this paper, we designed and developed Shepherd, a tool that enables
post-login measurements of unknown websites. As login processes are very
diverse, automatic logging in cannot achieve full coverage.
Moreover, the diversity of the employed login processes 
implies many design challenges. Previous efforts
using automated or semi-automated approaches managed up to
203 sites~\cite{RB14,MFK16}. In contrast, in our case study we
automatically logged into at least 6,273 sites.

We ran two experiments, one using credentials from an external source
(BugMeNot) and one using Facebook single sign-on. These experiments
found thousands of websites vulnerable to straightforward session
hijacking. However, as the comparison with the Facebook experiment
clearly showed, this cannot be extrapolated to other websites.
Nevertheless: equally clearly, the results of the BugMeNot experiment
cannot be ignored: out of the 6,273 BugMeNot sites where Shepherd could
successfully verify login automatically, 2,579 (41.4\%) are vulnerable.
It is an illusion to think that the problem of unprotected
authentication cookies only applies to sites for which BugMeNot has
credentials.

\subsection{Future work} 
First of all, we are working to support more single sign-on frameworks
and make these procedures more robust. This will allow us to get a
broader estimate of the number of sites susceptible to simple hijacking.
Secondly, we are planning to expand our study of session
security to account for more factors, such as session fixation or CSRF
resistance and cookie invalidation security. 

A followup work would be to combine logging in with the OpenWPM
framework to detect if the privacy of logged-in users is treated
equivalently to anonymous users or not.

\subsection{Acknowledgments}
We would like to thank the Technische Hochschule K\"oln for allowing us
to make use of their computing resources for this project. In addition,
we extend our thanks to Jun Pang, Olga Gadyatskaya, Agnieszka Jonker and
Jana Polednov\`a for translating keywords into Chinese, Russian, Polish,
French, Spanish, and Czech. Finally, we are grateful for comments we
received on earlier versions of this work from Marko van Eekelen, Greg
\'Alpar, the institutional review board of the Open University of the
Netherlands and the anonymous reviewers of ESORICS'18 and ICISC'18.

%
%
\bibliographystyle{alpha}
\bibliography{references}

%
%
\begin{subappendices}
	\renewcommand{\thesection}{\Alph{section}}%
\section{Login success rate compared with BugMeNot votes}
\label{sec:correlation_votes}
\setcounter{table}{0}
\renewcommand{\thetable}{A\arabic{table}}

User submitted credentials might be invalid or outdated. BugMeNot offers
to submit votes about the success of credentials. Separating invalid
from valid credentials would help to assess Shepherd's performance,
because this would eliminate the external cause for failed login
attempts. However, user submitted votes on its own need not be valid and
cannot fully be trusted -- yet they do provide some information. For
example, we found that a freshly added (invalid) credential receives a
100\% success score and one vote. Thus, when correlating votes in
BugMeNot with success rates of the Shepherd framework, we ignore
credentials with a 100\% score and less than 3 votes. We accumulated
success rates and vote counts for our dataset (see
Table~\ref{tbl:bmn_correlation}).

\begin{table}
	\centering
	\begin{tabular}{l|c|c|c|c} \multicolumn{3}{c|}{} & \multicolumn{2}{c}{Authenticated } \\ 
		Category & \hspace{0,1cm} Entire set \hspace{0,1cm}  &\hspace{0,1cm} Not authenticated 
		\hspace{0,1cm} & \hspace{0,1cm} includ. non verif. \hspace{0,1cm} & 
		\hspace{0,1cm} verified only\\
		\hline Domains  & 42,770 & 33,012  &8,805 & 5,066 \\
		~0 $\leq$ x $<$ ~30\%  & 14,327 & 12,581 & 1,208 & 408\\ 30 $\leq$ x $<$
		~70\% & 17,103 & 14,045 & 2,439 & 1,256\\ 70 $\leq$ x $\leq$ 100\%
		&11,340 & 6,386 &  4,558 & 3,402\\ \end{tabular} \vspace{1.5ex}
	\caption{Correlation of BugMeNot user votes and the result of scanning
		with Shepherd. All entries with less than 3 votes and 100\% were
		removed.} \label{tbl:bmn_correlation}
\end{table}

\begin{figure}	
	\centering
	\includegraphics[width=0.9\linewidth]{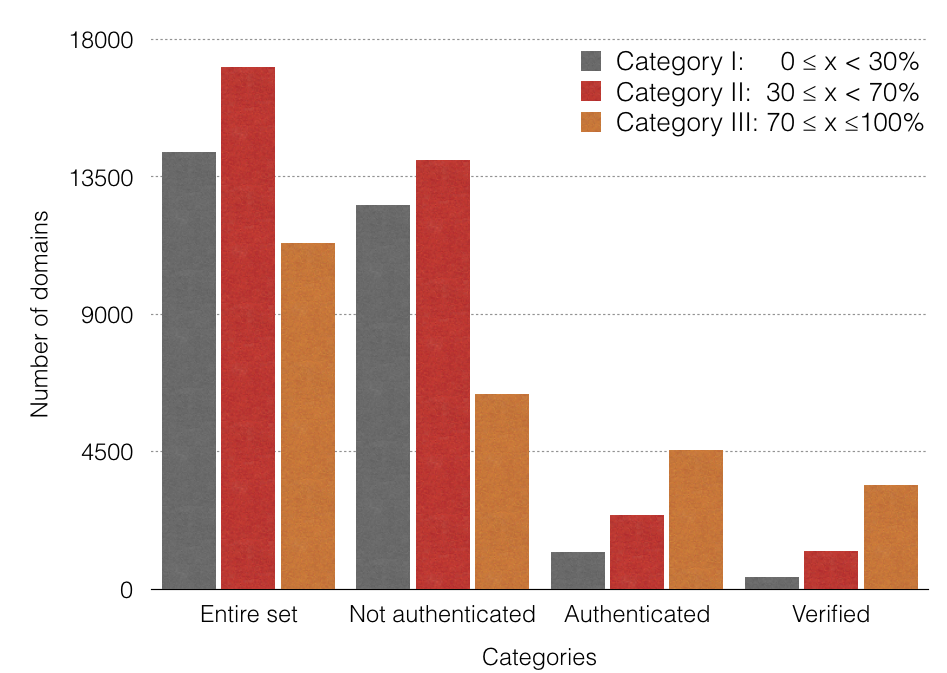}
	\caption{Correlation of BugMeNot user votes and Shepherd's results.}
	\label{fig:bmn_scanner_correlation}
\end{figure}

In some cases, BugMeNot provides multiple credentials for one domain. We
focus only on the credentials with the highest success rate. We sorted
these into 3 categories to determine the likelihood of successful login
(see Fig.~\ref{fig:bmn_scanner_correlation}). As shown in the figure,
there is a rough correlation between verified logins and highly rated
credentials, as well as the inverse: Shepherd fails in 97.1\% for lowly
rated credentials. 

\end{subappendices}
\end{document}